# Cost-Effective Optimization and Implementation of the CRT-Paillier Decryption Algorithm for Enhanced Performance

ZHENGWU HUANG*, DING DENG*, PENGYUE SUN, GUANGFU SUN, XIAOMEI TANG

College of Electronic Science and Technology, National University of Defense Technology; National Key Laboratory for Positioning, Navigation and Timing Technology, Changsha, China

To address the privacy protection problem in cloud computing, privacy enhancement techniques such as the Paillier additive homomorphism algorithm are receiving widespread attention. Paillier algorithm allows addition and scalar multiplication operations in dencrypted state, which can effectively protect privacy. However, its computational efficiency is limited by complex modulo operations due to the ciphertext expansion followed by encryption. To accelerate its decryption operation, the Chinese Remainder Theorem (CRT) is often used to optimize these modulo operations, which lengthens the decryption computation chain in turn. To address this issue, we propose an eCRT-Paillier decryption algorithm that shortens the decryption computation chain by combining precomputed parameters and eliminating extra judgment operations introduced by Montgomery modular multiplications. These two improvements reduce 50% modular multiplications and 60% judgment operations in the postprocessing of the CRT-Paillier decryption algorithm. Based on these improvements, we propose a highly parallel full-pipeline architecture to eliminate stalls caused by multiplier reuse in traditional modular exponentiation operations. This architecture also adopts some optimizations such as simplifying modular exponentiation units by dividing the exponent into segments and parallelizing data flow by multi-core instantiation. Finally, a high-throughput and efficient Paillier accelerator named MESA was implemented on the Xilinx Virtex-7 FPGA for evaluation, which can complete a decryption using 2048-bit key within 0.577ms under 100 MHz clock frequency. Compared to prior works, MESA demonstrates a throughput improvement of 1.16× to 313.21× under identical conditions, also with enhancements in area efficiency for LUT, DSP, and FF of 3.32× to 117.55×, 1.49× to 1.64×, and 2.94× to 9.94×, respectively.

**ACM Reference Format:**
First Author's Name, Initials, and Last Name, Second Author's Name, Initials, and Last Name, and Third Author's Name, Initials, and Last Name. 2018. The Title of the Paper: ACM Conference Proceedings Manuscript Submission Template: This is the subtitle of the paper, this document both explains and embodies the submission format for authors using Word. In Woodstock '18: ACM Symposium on Neural Gaze Detection, June 03–05, 2018, Woodstock, NY. ACM, New York, NY, USA, 10 pages. NOTE: This block will be automatically generated when manuscripts are processed after acceptance.

This work was supported in part by the National Natural Science Foundation of China (Grant No.62201585 and No. U20A20193), in part by the science and technology innovation Program of Hunan Province (Grant No.2023RC3004).

Authors'addresses:Z.W.Huang*, D.Deng*, P.Y.Sun, G.F.Sun and X.M.Tang, College of Electronic Science and Technology, National University of Defense Technology; National Key Laboratory for Positioning, Navigation and Timing Technology, Changsha 410073, China; emails: hzw@nudt.edu.cn, dengding15@nudt.edu.cn, sunnnpy@163.com, gfsunmail@163.com, txm_nnc@126.com

(*These authors contributed to the work equally and should be regarded as co-first authors. Corresponding author:P.Y.Sun)

## 1 INTRODUCTION

In recent years, the rapid development of cloud computing and Internet of Things (IoT) has provided strong support for big data applications such as outsourced computing [1], personal information retrieval [2] and distributed deep learning [3]. However, the data involved in these applications are often private and can leak personal privacy if leveraged by malicious server. Homomorphic encryption (HE) serves as an encryption scheme that establishes a one-to-one mapping between operations in the ciphertext domain and those in the plaintext domain, thereby enabling direct computation on ciphertext and thus preventing the leakage of plaintext data [4–6]. Therefore, homomorphic encryption is considered as one of the most promising techniques for privacy preservation.

Homomorphic encryption can be primarily categorized into Partially Homomorphic Encryption (PHE), Somewhat Homomorphic Encryption (SHE), and Fully Homomorphic Encryption (FHE). Among these, fully homomorphic encryption supports arbitrary computational operations on ciphertexts, but most of them also suffer serious ciphertext expansion and computational overhead problems [5,7,8]. In contrast, partially homomorphic encryption algorithms such as the Paillier algorithm can support infinite additive homomorphic operations with better computational efficiency. Currently, the Paillier algorithm has been widely adopted in various fields such as secure multi-party computation, threshold signatures, and electronic voting. Nevertheless, its computational speed remains one of the main factors limiting its practical application. Especially during its decryption operation, the width of the operands is double that of the plaintext, causing significant increase in computational load. Therefore, optimizing and accelerating the decryption operation is crucial for the promotion of the Paillier algorithm.

From the perspective of algorithm, Chinese Remainder Theorem (CRT) is a widely used optimization approach due to its property of decomposing modulus values [9]. To accelerate the decryption operation, El Makkaoui et al. [10] proposed a Cloud-Paillier fast decryption algorithm (CRT-base Paillier) based on CRT in 2018 and adopted Bézout's lemma to optimize the inverse CRT process. Based on this, they proposed a multi-prime modulus Cloud-Paillier decryption scheme [11], which uses CRT to split the modulus into multiple prime numbers, achieving further acceleration of decryption. In the same year, Ogunseyi et al. [12] introduced a variable $k$ for the parameter-precomputation process of decryption based on the work in literature [10]. They reorganized and optimized the $k$, $L$ function, and CRT process to reduce the computational overhead of precomputation, but did not optimize the hardware implementation. Even worse, the number of moduli involved in decryptions increases from two to five after its optimization, which lengthens the hardware logic chain and increases the complexity of controlling circuits.

From the perspective of hardware, architecture modification is another effective approach to accelerate Paillier. In 2016, San et al. designed a high-speed FPGA-based co-processor architecture to perform Paillier homomorphism operations [13], which employs high-radix Montgomery modular algorithm and pipeline multiplier to accelerate the calculation. In 2019, Cai et al. [14] designed a dedicated Paillier hardware architecture for ASIC, exploiting the parallelism of the high-radix arithmetic operations and decryption. Whereas, the works in [13, 14] are both single-core designs, which use a single modular multiplication unit to perform modular multiplications and modular exponentiation operations. Although this method reduces hardware overhead, it results in stalls during modular exponentiation. Moreover, those designs lack fine-grained computation partition, leaving significant room of area optimization for multi-core designs. In 2020, Bahadori et al. [15] proposed a multi-core Paillier hardware accelerator based on microcode. Similarly, this design also has the same drawbacks as the single-core design mentioned above. In 2024, Che et al. [27] proposed a heterogeneous Paillier accelerator design for federated learning, incorporating multiple processing elements to achieve Single Instruction Multiple Data (SIMD) operations for large numbers. Whereas this design only introduces pipelines into the modular multipliers of the processing elements and assigns individual task scheduler to each processing element, the resource overhead and throughput are suboptimal.

Furthermore, as the bottleneck of the decryption operation, the modular exponentiation is also an exploitable optimization for Paillier acceleration. Many researchers use the Montgomery modular multiplication (MM) algorithm [16–19] to accelerate modular multiplication, which is a key component of modular exponentiation. To pursue high performance, some works removed the final judgment steps in the high-radix Montgomery modular multiplication algorithm [17,19,20], which may lead to the result of modular multiplier one modulus larger than the theoretical value. Although this does not affect intermediate computations during successive modular multiplications, it requires an extra step of removing the excess modulus during modulus switching or operation switching, which can lead to large area and latency overhead, especially for its CRT version.

To avoid the above disadvantages, without compromising algorithm security or hardware feasibility, this paper presents two methods to shorten the computation chain in CRT-base Paillier decryption. Based on these methods, we propose an improved eCRT-Paillier decryption algorithm. Thus, the decryption computation chain is significantly shortened and the control logic is simplified. Additionally, we propose a full-pipeline architecture with segment organization of modular exponentiation units. Main contributions are highlighted as follows:

1) An efficient eCRT-Paillier algorithm based on CRT-base Paillier is proposed by combining precomputed parameters to shorten the computation chain. Concretely, by combining the precomputed parameter of the two modular multiplications after modular exponentiation, two modular multiplications are reduced to one, thus shortening the computation chain. Besides, the correctness of the combination of parameters is proved.
2) The judgment operation, which is used to ensure the correctness of Montgomery modular multiplications, is proved removable in our eCRT-Paillier algorithm. We show that only a single judgment operation is required after the final modular multiplication operation through formula derivation. This method further reduces the complexity of hardware control logic and saves area overhead.
3) Based on the proposed eCRT-Paillier algorithm, a highly parallelized and configurable full-pipeline architecture for Paillier acceleration is proposed. Following this architecture, a Paillier accelerator named MESA is implemented. By meticulous timing partition, the architecture divides the eCRT-Paillier process into three relatively independent phases: preprocessing, modular exponentiation, and postprocessing. Correspondingly, a highly parallel pipeline structure is designed to eliminate the stalls in modular exponentiation caused by the reuse of modular multipliers. In addition, the exponent is divided into segments to simplify the data flow scheduling of the multi-core modular exponentiation units, which helps to achieve high throughput with small area overhead.

The rest of this paper is organized as follows. Section II introduces the symbols and algorithms, including traditional Paillier decryption and CRT-Paillier decryption (CRT-base), the fast modular exponentiation and modular multiplication. Section III introduces eCRT-Paillier decryption algorithm as an improved version based on CRT-base Paillier. Section IV provides details of the multi-core Paillier acceleration architecture and the MESA accelerator proposed for high-throughput applications. Section V evaluates the proposals through hardware implementation on an FPGA and makes a comparison with other works. Finally, this paper is concluded in Section VI.

## 2 BACKGROUND

### 2.1 Theoretical Foundations

Let $Z_n^*$ denote the set of positive integers modulo $n$, and its complete residual class be $Z_n = \{0,1,\cdots,n-1\}$. The bit width of the public key $n$ is denoted as $N$, serving as the reference for describing the bit widths of other parameters in Paillier.

The plaintext is denoted by *m*, and the ciphertext is denoted by *c*. The Montgomery domain for mod *n* is denoted as $R_0$ and for mod $n^2$ is denoted as $R_1$. Montgomery modular multiplication is denoted as MM, and modular exponentiation is denoted as ME. The greatest common divisor and the least common multiple are respectively denoted as gcd and lcm.

**Definition 1**: The function *L* is defined over $S = \{x < n^2 \mid x \bmod n = 1\}$ by: $L_n(x) = \dfrac{x-1}{n}$

**Theorem 1**: [Chinese Remainder Theorem] The Chinese Remainder Theorem provides a solution for the following system of linear congruences with one unknown:

$$\begin{cases} x \equiv a_1 \pmod{p_1} \\ x \equiv a_2 \pmod{p_2} \\ \quad \vdots \\ x \equiv a_n \pmod{p_n} \end{cases}$$

The theorem states that when $p_1, p_2, \ldots p_n$ are co-prime, for any $a_1, a_2, \ldots a_n$, the above equations have a solution, and a general solution can be constructed by the following steps:

**Step 1:** Calculate $M = \prod_{i=1}^{n} p_i, M_i = M / p_i$.

**Step 2:** And then calculate the inverse $t_i$ of $M_i$ module $p_i$, that is:

$$t_i = M_i^{-1} \bmod p_i$$

**Step 3:** Finally, we can get the general solution *x*:

$$x = \sum_{i=1}^{n} a_i t_i M_i \bmod M$$

## 2.2 Traditional Paillier Decryption Algorithm

In the Paillier cryptosystem, the public key consists of *n*、*g*, and the private key consists of $\lambda$、$\mu$. The algorithm parameters *p*, *q* and *g* are chosen such that *p* and *q* are co-primes with equal bit lengths, satisfying $\gcd(pq, (p-1)(q-1)) = 1$. The value $\lambda$ is set to $\text{lcm}(p-1, q-1)$, and *n* is the product of *p* and *q* (i.e., $n = pq$). The parameter *g* must satisfy $g \in Z_{n^2}^*$ and $\gcd(L(g^\lambda \bmod n^2), n) = 1$. The private key component $\mu$ is computed as $\left(L(g^\lambda \bmod n^2)\right)^{-1} \bmod n$. The Paillier cryptosystem consists of three main parts: encryption, evaluation, and decryption. This subsection focuses on the decryption process. For ciphertext $c \in Z_{n^2}^*$, the decryption proceeds as follows:

$$m = Dec(c) = L(c^\lambda \bmod n^2) \cdot \mu \bmod n \quad (1)$$

## 2.3 CRT-base Paillier Decryption Algorithm

As shown in Equation (1), the traditional Paillier algorithm performs decryption over $Z_n^*$. To ensure sufficient security, the bit width of the modulus *n* is typically as large as several thousand bits, resulting in highly time-consuming modular exponentiation operations. This becomes the critical bottleneck that limits the decryption speed of Paillier. To improve the efficiency of the algorithm, the CRT can be used to decompose the modulo operation in Paillier algorithm, and the modulo $n^2$ operation can be converted into modulo $p^2$ and modulo $q^2$ operations, so as to reduce the complexity of modular exponentiation operation.

By applying Theorem 1 to equation (1), we can obtain the CRT-base Paillier decryption algorithm, and the specific algorithm steps are as follows:

**Step 1**: Precompute the modular inverses $w_3$ and $w_4$ of *p* and *q*, with $w_3 = q^{-1} \bmod p$, $w_3 = p^{-1} \bmod q$. Let $l_3 = w_3 \cdot q$,

$l_4 = w_4 \cdot p$ , precompute $e_p, e_q$.

$$e_p = \left(L_p(g^{p-1} \bmod p^2)\right)^{-1} \bmod p \quad (2)$$
$$e_q = \left(L_q(g^{q-1} \bmod q^2)\right)^{-1} \bmod q$$

**Step 2**: And then calculate $m_p$ and $m_q$ as,

$$m_p = \left(L_p\left(c^{p-1} \bmod p^2\right)\right)e_p \bmod p \quad (3)$$
$$m_q = \left(L_q\left(c^{q-1} \bmod q^2\right)\right)e_q \bmod q$$

**Step 3:** Finally, by appyling CRT, we can get plaintext *m*.

$$m = \left(m_p \cdot l_3 \bmod n + m_q \cdot l_4 \bmod n\right) \bmod n \quad (4)$$

### 2.4 Fast Modular Exponentiation Algorithm

In Paillier algorithm, modular exponentiation consumes the most computation time. As one of the most efficient algorithms for performing modular exponents, fast modular exponents can reduce the complexity of algorithms from $O(2^N)$ to $O(N)$. The algorithm also requires only $2N$ modular multiplications in the worst case. The fast modular exponentiation algorithm is categorized into L-R and R-L binary modular exponentiation algorithms according to the different scanning order [21]. These two modular exponentiation algorithms transform the modular exponentiation operation into an iterative operation of modular multiplication and modular square through binary expansion of the exponentiation. Specifically, when the exponential bit of the scan is 1, the modular square and modular multiplication operations are performed simultaneously, and when the exponential bit of the scan is 0, only the modular square operation is performed. However, this characteristic where the number of operations varies with the scanned exponent bit, can lead to data correlation between the two algorithms in terms of time and power consumption, which can easily leak key information about the exponent bits.

To eliminate the data correlation between side-channel information, such as power consumption and time, and the exponent, Joye et al. introduced redundant operations in L-R algorithms [22], proposing the Montgomery power ladder algorithm shown in Algorithm 1. In this algorithm, regardless of whether the exponent bit is 0 or 1, both modular multiplication and modular square operations are performed, thereby avoiding the leakage of key information through obvious side-channel information.

---

**AlGORITHM1:** Montgomery Power Ladder Method

**Input:** $S = \text{MM}(1, R^2, p), Z = \text{MM}(a, R^2, p)$ , $p, b = (b_{N-1}, b_{N-2}, \cdots, b_0) = \sum_0^{N-1} b_i 2^i$

**output:** $a^b \bmod p$

1: **for**(*i=N*-1;*i*≥ 0;*i=i*-1) **do**
2:   **if**($b_i$==1) **then**
3:      $S = \text{MM}(S, Z, p)$
4:      $Z = \text{MM}(Z, Z, p)$
5:   **else**
6:      $Z = \text{MM}(S, Z, p)$
7:      $S = \text{MM}(S, S, p)$
8:   **endif**

---

9: **endfor**
10: $S = \text{MM}(1, S, p)$
11: **return** $S$

## 2.5 CIOS Montgomery Modular Multiplication Algorithm

The basic operator for modular exponentiation is modular multiplication. The Montgomery modular multiplication algorithm [23] is widely adopted to accelerate large-number modular multiplication. The core idea of this algorithm is to use shift, modular addition and multiplication to replace division, thus reducing computational complexity. Depending on operand segmentation granularity, Montgomery modular multiplication schemes can be categorized into radix-2 [24] (denoted as $MM_B$) and high-radix (denoted as $MM_H$) types. The radix-2 scheme processes data bit-by-bit, implementing multiplication operations by AND gates, resulting in smaller area overhead. However, due to its bit-by-bit scanning approach, its performance is poor, typically requiring $\lceil \log_2 p + 2 \rceil$ cycles to complete computation, where $p$ is the modulus. In contrast, high-radix schemes support processing multiple bits of operands simultaneously, significantly improving computational speed.

In high-radix schemes, the Montgomery modular multiplication algorithm based on Coarsely Integrated Operand Scanning (CIOS) [25] is widely adopted due to its feature where data for all multiplication operations arrive simultaneously, which facilitates efficient implementation on DSPs (Digital Signal Processors). The specific operational steps are shown in Algorithm 2.

---

**AlGORITHM2:** CIOS Montgomery Modular Multiplication

**Input:** $a, b, p, R = 2^{\left(\lceil \lceil \log_2 p + 2 \rceil / \textbf{word size} \rceil \cdot \textbf{word size}\right)}, w = 2^{\textbf{word size}}, p' = -p^{-1} \bmod R, l = \lceil \log_2 (R/w) \rceil$

**Output:** $abR^{-1} \bmod p$

**Initial:** $t = 0$

1: **for** $i = 0$ *to* $l-1$ **do**
2: $(c, s[0]) = a[0] \cdot b[i] + t[0]$
3: **for** $j = 1$ *to* $l-1$ **do**
4: $(c, s[j]) = a[j] \cdot b[i] + t[j] + c$
5: **end for**
6: $s[j+1] = c$
7: $m = s[0] \cdot p' \bmod w$
8: $(c, null) = s[0] + m \cdot p[0]$
9: **for** $j = 1$ *to* $l-1$ **do**
10: $(c, t[j-1]) = s[j] + m \cdot p[j] + c$
11: **end for**
12: $t[j] = s[j+1] + c$
13: **end for**
14: **return** $t$

---

## 2.6 CRT-Paillier with Montgomery Algorithm

Algorithm 3 represents the complete process of CRT-base Paillier decryption using the Montgomery modular multiplication algorithm and the Montgomery power ladder algorithm. To facilitate subsequent descriptions, this algorithm is divided into four parts: precomputation, preprocessing, modular exponentiation, and postprocessing. In the algorithm, the moduli $p^2$ and $q^2$ correspond to the Montgomery domain and are consistent with the modulus $n$ ($p^2$, $q^2$, and $n$ have the same bit width), while the moduli $p$ and $q$ correspond to the Montgomery domain as $R_2$. The postprocessing phase includes two sections marked with purple (Object 1) and orange boxes (Object 2), which are the targets of the optimization proposed. Object 1 integrates precomputed parameters for optimization, while Object 2 eliminates redundant elements. The following sections will provide a detailed explanation of these two optimization techniques.

---

**AlGORITHM3: CRT-base Paillier**

**Input:** $c$  **Output:** $m$

**Precompute:** $y_p = R_1^2 \bmod p^2$, $y_q = R_1^2 \bmod q^2$
$l_{3,R} = l_3 \cdot R_0 \bmod n$, $l_{4,R} = l_4 \cdot R_0 \bmod n$
$e_{p,R} = e_p \cdot R_2 \bmod p$, $e_{q,R} = e_q \cdot R_2 \bmod q$

**Preprocessing:**
1: $S_p = \text{MM}_\text{B}(c, y_p, p^2) = cy_p R_1^{-1} \bmod p^2 = cR_1 \bmod p^2$
$S_q = \text{MM}_\text{B}(c, y_q, q^2) = cy_q R_1^{-1} \bmod q^2 = cR_1 \bmod q^2$

**Modular Exponentiation:**
2: $c_{e,p} = \text{ME}(S_p, p-1, p^2)$, $c_{e,q} = \text{ME}(S_q, q-1, q^2)$

**Postprocessing:**
3: Execution judgment: **[object 2]**
$U_p = (c_{e,p} \geq p^2)\ ?\ c_{e,p} - p^2 : c_{e,p}$
$U_q = (c_{e,q} \geq q^2)\ ?\ c_{e,q} - q^2 : c_{e,q}$

4: Compute L function:
$L_{c,p} = L_p(U_p)$, $L_{c,q} = L_q(U_q)$

5: Execution MM:
$m_p = \text{MM}_\text{B}(L_{c,p}, e_{p,R}, p)$, $m_q = \text{MM}_\text{B}(L_{c,q}, e_{q,R}, p)$

6: Execution judgment:
$m_p = (m_p \geq p)\ ?\ m_p - p : m_p$
$m_q = (m_q \geq p)\ ?\ m_q - p : m_q$

7: Execution MM: **[object 1]**
$M_p = \text{MM}_\text{B}(m_p, l_{3,R}, n)$, $M_q = \text{MM}_\text{B}(m_q, l_{4,R}, n)$

8: Execution judgment:
$M_p = (M_p \geq n)\ ?\ M_p - n : M_p$
$M_q = (M_q \geq n)\ ?\ M_q - n : M_q$

9: Execution ModAdd:
$m = \text{ModAdd}(M_p, M_q, n)$

---

## 3 ALGORITHM OPTIMIZATION

### 3.1 Shorten the Decryption Chain

Compared to the traditional Paillier decryption algorithm, the CRT-base decryption algorithm adds one modular multiplication with precomputed parameters in the post-processing phase, as shown in Step 7 of Algorithm 3 (corresponding to object 1). Steps 5 and 7 correspond to Equations (3) and (4), respectively. It is observed that parameters $e_p, l_3, e_q, l_4$ in Equations (3) and (4) are all obtained through precomputation. To reduce the number of modular multiplications in the postprocessing phase, the precomputed parameters $e_p$ and $l_3$, and $e_q$ and $l_4$, are integrated, resulting in the merged precomputed parameters $t_p$ and $t_q$. This successfully shifts the additional modular multiplication in the postprocessing step of the CRT-base decryption algorithm to the precomputation phase, thus decreasing the number of modular multiplications in the postprocessing phase.

Below is the derivation process for integrating parameters $e_p$ and $l_3$ (or $e_q$ and $l_4$) into a single parameter. Equation (5) is obtained by substituting the precomputed parameters into Equation (4).

$$m = \left[\left\{L_p\left(c^{p-1} \bmod p^2\right) \cdot e_p \bmod p\right\} \cdot \left((q^{-1} \bmod p)q\right) \bmod n + \cdots\right] \bmod n \quad (5)$$

Let $r_x = L_p\left(c^{p-1} \bmod p^2\right)$, $r_y = e_p$, $r_z = q^{-1} \bmod p$. Where $t \geq 0$, and substitute into Equation (5) to obtain Equation (6).

$$\begin{aligned}
m &= \left[\left(\left(r_x r_y \bmod p\right) \cdot r_z \cdot q\right) \bmod n + \cdots\right] \bmod n \\
&= \left[\left((p \cdot t + r_{xy}) \bmod p\right) \cdot r_z \cdot q \bmod n + \cdots\right] \bmod n \\
&= \left[\left(r_{xy} \bmod p\right) \cdot r_z \cdot q \bmod n + \cdots\right] \bmod n \\
&= \left[r_{xy} \cdot r_z \cdot q \bmod n + \cdots\right] \bmod n
\end{aligned} \quad (6)$$

Given that $n = pq$, by the properties of modular congruence, we obtain Equation (7):

$$\begin{aligned}
m &= \left[\left(r_{xy} \cdot r_z \cdot q + p \cdot t \cdot r_z \cdot q\right) \bmod n + \cdots\right] \bmod n \\
&= \left[\left(p \cdot t + r_{xy}\right) \cdot r_z \cdot q \bmod n + \cdots\right] \bmod n \\
&= \left[\left(r_x r_y \cdot r_z \cdot q\right) \bmod n + \cdots\right] \bmod n \\
&= \left[\left(L_p\left(c^{p-1} \bmod p^2\right) \cdot e_p \cdot l_3 \bmod n\right) \bmod n + \cdots\right] \bmod n \\
&= \left[\left(L_p\left(c^{p-1} \bmod p^2\right) \cdot (e_p \cdot l_3 \bmod n)\right) \bmod n + \cdots\right] \bmod n
\end{aligned} \quad (7)$$

All transformations from Equation (5) to Equation (7) are equivalent. Therefore, after combining the precomputed parameters following the above method, we can let $t_p = e_p \cdot l_3 \bmod n$ and $t_q = e_q \cdot l_4 \bmod n$. Substituting $t_p, t_q$ into Equation (7), the decryption calculation formula can ultimately be rewritten as Equation (8):

$$m = \left[L_p\left(c^{p-1} \bmod p^2\right) \cdot t_p \bmod n + L_q\left(c^{q-1} \bmod q^2\right) \cdot t_q \bmod n\right] \bmod n \quad (8)$$

## 3.2 Remove Judgment Operation from Postprocessing

Modular multiplication, as a subprocess of modular exponentiation, is typically accelerated using the Montgomery algorithm. To perform modular exponentiation more efficiently, Li et al.[17] removed the final judgment step in the Montgomery modular multiplication algorithm, requiring only one judgment at the end of the modular exponentiation, thus saving a significant amount of time on comparisons and subtractions. However, when operations involve switching modulus, it is necessary to determine whether the modular multiplication result exceeds the modulus. After optimizing decryption with the CRT, the number of modulus types increases from 2 to 5. Specifically, in CRT-base Paillier, there are 5 types of moduli involved: $p, q, p^2, q^2, n$. Even in our simplified decryption algorithm (Equation (8)), which includes 3 types of moduli ($p^2, q^2, n$), the number of modulus switches increases from two to four. If this approach is directly applied to compute Equation (8), additional judgment steps will be introduced during postprocessing, such as Step 3 in Algorithm 3 (object 1) and Step 8 (Step 6 is removed due to the consolidation of Steps 5 and 7). This makes the computation chain quite lengthy and increases the complexity of control logic design and hardware overhead.

To further shorten the computation chain and simplify control logic while lowering resource overhead, we conducted theoretical derivations on the algorithm and proposed a method that saves both time and area overhead without affecting the correctness of the results: specifically, by directly removing Step 3 from Algorithm 3. The detailed proof process is as follows:

(1) For cases where the result does not exceed the modulus, it is obvious that the step can be directly removed.

(2) For cases where the result exceeds the modulus, let:

$$L_{c,p} = \frac{c_{e,p} + p^2 - 1}{p} = \frac{c_{e,p} - 1}{p} + p \quad (9)$$

Substituting Equation (9) into Equation (8), we obtain:

$$\begin{aligned}
m_p &= L_{c,p} \cdot t_p \bmod n \\
&= \left(\frac{c_{e,p}-1}{p} + p\right) \cdot \left((e_p \cdot (q^{-1} \bmod p) \cdot q) \bmod n\right) \bmod n \\
&= \left[\frac{c_{e,p}-1}{p} \cdot \left((e_p \cdot (q^{-1} \bmod p) \cdot q) \bmod n\right) + \left(p \cdot e_p \cdot (q^{-1} \bmod p) \cdot q\right) \bmod n\right] \bmod n \quad (10)\\
&= \left[\frac{c_{e,p}-1}{p} \cdot \left((e_p \cdot (q^{-1} \bmod p) \cdot q) \bmod n\right) + \left(e_p \cdot (q^{-1} \bmod p) \cdot n\right) \bmod n\right] \bmod n \\
&= \left[\frac{c_{e,p}-1}{p} \cdot \left((e_p \cdot (q^{-1} \bmod p) \cdot q) \bmod n\right)\right] \bmod n
\end{aligned}$$

According to Equation (10), the excess modulus value $p$ will be automatically reduced in subsequent modular operations, so no judgment step is necessary and it can be directly removed.

### 3.3 Algorithm Discussion

The complete process of eCRT-Paillier decryption, combining the Montgomery modular multiplication algorithm and the Montgomery power ladder algorithm, is shown in Algorithm 4. Compared to CRT-base Paillier, the eCRT-Paillier not only simplifies the algorithm flow but also decreases the number of modular multiplications and judgments required in postprocessing.

---

**AlGORITHM 4: eCRT-Paillier**

---

**Input:** $c$ **Output:**

**Precompute:** $y_p = R_1^2 \bmod p^2$, $y_q = R_1^2 \bmod q^2$

$t_{p,R} = t_p \cdot R_0 \bmod n$, $t_{q,R} = e_q \cdot R_0 \bmod n$

**Preprocessing:**

1: $S_p = \mathrm{MM_B}(c, y_p, p^2) = c y_p R_1^{-1} \bmod p^2 = c R_1 \bmod p^2$

$S_q = \mathrm{MM_B}(c, y_q, q^2) = c y_q R_1^{-1} \bmod q^2 = c R_1 \bmod q^2$

**Modular Exponentiation:**

2: $c_{e,p} = \mathrm{ME}(S_p, p-1, p^2)$, $c_{e,q} = \mathrm{ME}(S_q, q-1, q^2)$

**Postprocessing:**

3: Compute L function:

$L_{c,p} = L_p(c_{e,p})$, $L_{c,q} = L_q(c_{e,q})$

4: Execution MM:

---

$$m_p = \mathrm{MM_B}(L_{c,p}, t_{p,R}, n)$$
$$m_q = \mathrm{MM_B}(L_{c,q}, t_{q,R}, n)$$

5: Execution judgment:
$$m_p = (m_p \geq n) \ ? \ m_p - n : m_p$$
$$m_q = (m_q \geq n) \ ? \ m_q - n : m_q$$

6: Execution ModAdd:
$$m = \mathrm{ModAdd}(m_p, m_q, n)$$

For the Paillier decryption algorithm with a public key width of $N$, the bit widths of the moduli $p, q, p^2, q^2$ and $n$ are $N/2, N/2, N, N$ and $N$, respectively. Considering Montgomery domain conversion and hardware resource reuse, operations with different bit widths are converted to 2$N$-bit width operations, with the conversion rule that a 2$N$-bit operation is equivalent to two $N$-bit operations. For CRT-base Paillier, the number of modular multiplications required in postprocessing is $2_N + 2_N = 2_{2N}$ (where the subscript denotes the bit width), and the number of judgments required is $2_N + 2_{N/2} + 2_N = 2.5_{2N}$. For the improved eCRT, the number of modular multiplications required in postprocessing is $2_N = 1_{2N}$, and the number of judgments required is $2_N = 1_{2N}$. The results of the algorithm comparison are shown in Table I.

Table I. Comparison of Postprocessing Modulo and Judgment Counts of Different Decryption Algorithms

| algorithm | modular multiplications | judgments |
|---|---|---|
| CRT-base | 2 | 2.5 |
| eCRT | 1 | 1 |
| **Reduction*** | **50%** | **60%** |

＊：(1−CRT-base/eCRT)×100%

This paper introduces an improvement to the postprocessing phase of the CRT-base Paillier algorithm. By combining precomputed parameters and removing redundant judgment steps, the length of the computation chain in postprocessing is significantly shortened. Compared with CRT-base Paillier, the improved eCRT algorithm achieves a 50% reduction in modular multiplications and a 60% reduction in judgments.

## 4 HARDWARE ARCHITECTURE

### 4.1 Overall Design

The proposed full-pipeline architecture employs a reconfigurable design, supporting compile-time configurability. This architecture allows for customization of modulus bit-width and the number of pipeline stages in modular exponentiation units according to different security requirements. Fig. 1 illustrates the Paillier accelerator MESA implemented based on this architecture, employing a 3-stage modular exponentiation pipeline (including preprocessing and postprocessing, totaling 5 stages). The accelerator consists of processing elements (PEs), a data path, a configuration unit (Cfg_unit), and a control unit.

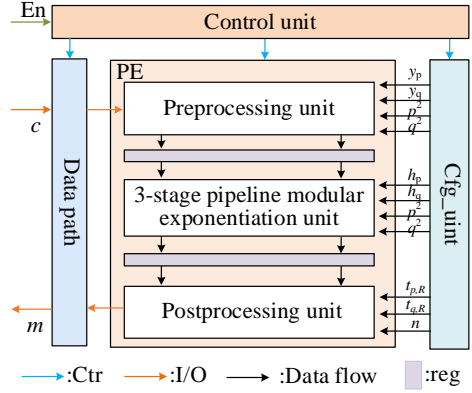

**Fig. 1 .** Top-level architecture of the MESA

The PE units, as the core processing units of MESA, are used to compute the eCRT-Paillier decryption operation as shown in Algorithm 4. The data path is responsible for the input of the ciphertext $c$ and the output of the plaintext $m$. The Cfg_unit, acting as a parameter configuration unit, provides precomputed parameters and corresponding moduli to the PE units, primarily implemented via RAM. The control unit performs data path read/write control, enables and resets the PE units, and generates read addresses and read control for the Cfg_unit.

The operation principle of MESA involves accepting the ciphertext $c$ and enabling signal En as inputs. After being enabled, the control unit sequentially loads the corresponding precomputed parameters from the Cfg_unit, according to the five pipeline stages of the PE, enabling the decryption of the input ciphertext $c$ in a pipelined manner.

**4.2  PE Unit**

To simplify the data flow scheduling within the PE unit, we employ an exponential segmentation method to organize multiple modular exponentiation units within the unit. To fully leverage the parallel capabilities of the eCRT decryption algorithm, the PE is implemented with two completely symmetric branches, $p$ and $q$. As shown in Fig. 2, the PE unit comprises a preprocessing unit, a 3-stage pipelined modular exponentiation unit, and a postprocessing unit. Each component serves specific functions, detailed as follows:

1) **Preprocessing Unit:** Composed of two $MM_B$ modules (radix-2 Montgomery modular multipliers), this unit is responsible for converting the input ciphertext $c$ into the Montgomery domain.
2) **3-Stage Pipelined Modular Exponentiation Unit:** This unit is structured with six ME modules organized through exponential segmentation, implementing a pipelined approach for modular exponentiation. This design significantly simplifies the data flow management among the ME modules. The ME unit implements the functionality of Algorithm 1, with the first two stages of ME only executing lines 1-9 of Algorithm 1, i.e., keeping the operands in the Montgomery domain, while the last stage of ME executes the complete algorithm.
3) **Postprocessing Unit:** This section includes two div modules (large integer dividers), two $MM_B$ modules, one add module (adder), and three com/sub modules (comparators and subtractors). The div modules are used to implement the $L$ function during decryption. Given the implementation of the restoring division algorithm by Hiasat et al. [26], the subtraction by 1 operation within the $L$ function is eliminated [13]. The modular multipliers are utilized to perform step 4 of Algorithm 4. The first com/sub modules are used to determine if the modular multiplication result exceeds the modulus. The subsequent add and com/sub modules handle the modular addition operations. To minimize hardware resource consumption, the modular multipliers in this unit are shared with the preprocessing unit.

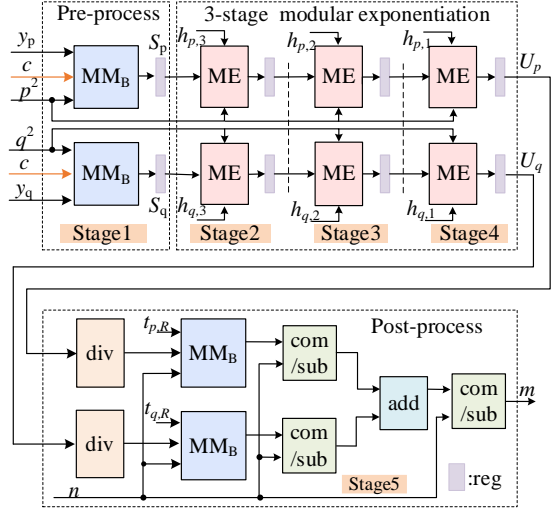

**Fig. 2 .** PE unit

The operational principle of the PE unit is as follows: Initially, the cfg_unit loads precomputed parameters $y_p = R_1^2 \bmod p^2$, $y_q = R_1^2 \bmod q^2$, and $p^2$, $q^2$, where $R_1 = R^2$. The preprocessing unit performs modular multiplication to obtain $S_p$ and $S_q$ respectively. Subsequently, the exponents are segmented and loaded sequentially from the cfg_unit into the ME modules, adhering to the segmentation relationship: $p-1 = \{h_{p,3}, h_{p,2}, h_{p,1}\}$, $q-1 = \{h_{q,3}, h_{q,2}, h_{q,1}\}$, where curly braces denote bit concatenation. After passing through the 3-stage pipelined modular exponentiation unit, the outputs from $S_p$ and $S_q$ are $U_p = c^{p-1} \bmod p^2$ and $U_q = c^{q-1} \bmod q^2$, respectively. During postprocessing, the div modules are first used to complete the computation of the $L$ function in the equation, yielding $L_q(c^{q-1} \bmod q^2)$ and $L_p(c^{p-1} \bmod p^2)$. Finally, the precomputed parameters $t_{p,R}$ and $t_{q,R}$, as well as the modulus $n$, are loaded to perform modular multiplication, judgment, and modular addition operations, resulting in the plaintext $m$. The detailed data flow is illustrated in Fig. 3.

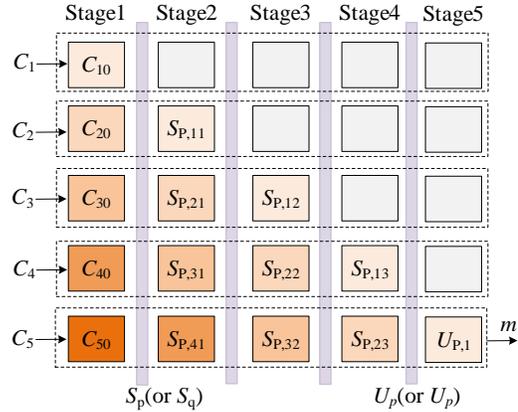

**Fig. 3.** Decryption Data Flow

In Fig. 3, $c_{i,0}$ represents the ciphertext $c_i$ undergoing Montgomery domain conversion. $S_{p,ij}$ (or $S_{q,ij}$) represents the i-th ciphertext $c_i$ undergoing modular exponentiation at the j-th stage, specifically $S_{p,i(j-1)}^{h_{p,j}} \mod p^2$ (or $S_{q,i(j-1)}^{h_{q,j}} \mod q^2$). The ciphertext $c_i$ after undergoing a five-stage pipeline, completes the decryption operation to yield the plaintext $m_i$.

### 4.3 ME Unit

Fig. 4 illustrates the internal structure of the ME designed in this paper. The ME is mainly composed of a finite state machine, two high-radix modular multipliers (MM$_H$), three multiplexers (MUX), and a shift register. The finite state machine controls the MUX switching, and enables the MM$_H$ and the shift register, while the shift register performs the shifting operation of the exponent $b$, and the MM$_H$ is used to complete the modular multiplication and modular square operations in Algorithm 1. Among them, the MM$_H$ implemented the CIOS high-radix Montgomery modular multiplication algorithm, and to more efficiently perform modular multiplication, this unit uses the DSP48E1 block for acceleration.

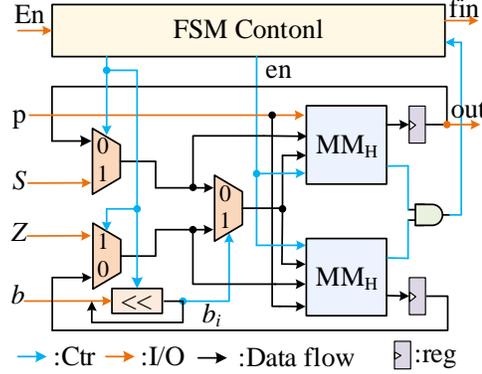

**Fig. 4.** ME unit

The ME takes the Montgomery domain conversion parameters $S$, the Montgomery domain representation of the base $a$ (denoted as $Z$), the exponent $b$, and the modulus $p$ as inputs. Under the control of the FSM, it repeatedly performs modular multiplication and modular square operations. Upon completion, it outputs the calculation result and a flag. The iterative process is shown in Fig. 5. It can be observed that as the exponent $b_i$ alternates between 0 and 1, the modular multiplication and modular square operations are correspondingly executed alternately in the upper and lower modular multipliers. Therefore, regardless of the value $b_i$, the power consumption information of the ME remains identical.

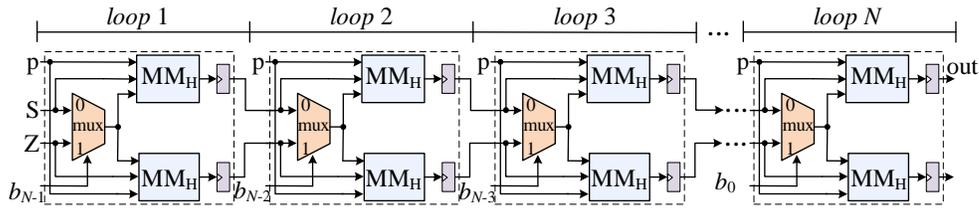

**Fig. 5.** Iterative process of ME

### 4.4 Highly Parallel Pipelined Design

MESA adopts a full-pipeline structure. Through fine-grained division and organization of the decryption operation logic chain, the most time-consuming modular exponentiation operation is designed with a pipeline, eliminating idle waiting between adjacent modular exponentiation operations and thus achieving 100% utilization of the modular exponentiation unit (ME). In the following, we will analyze and compare the traditional multi-core architecture with the proposed highly parallel fully pipelined structure.

In previous hardware acceleration designs for Paillier [13–15], pipelining acceleration was typically applied only to the modular multiplication operator, lacking a detailed pipelined design for decryption operations. As shown in Fig. 6(a), the traditional multi-core architecture reused $MM_H$ in both modular exponentiation and modular multiplication to reduce hardware overhead, in scheduling operators such as MM and $L(x)$. Although this design reduces the hardware overhead of a single operation core, modular exponentiation will be stalled during postprocessing modular multiplication operation, thereby reducing decryption efficiency. Additionally, when multi-core is invoked, non-modular multiplication operators in each operation core remain idle for more than 95% of the time, reducing hardware utilization. To address this, we propose a highly parallel pipelined structure, based on the eCRT algorithm, to eliminate stalls in modular exponentiation.

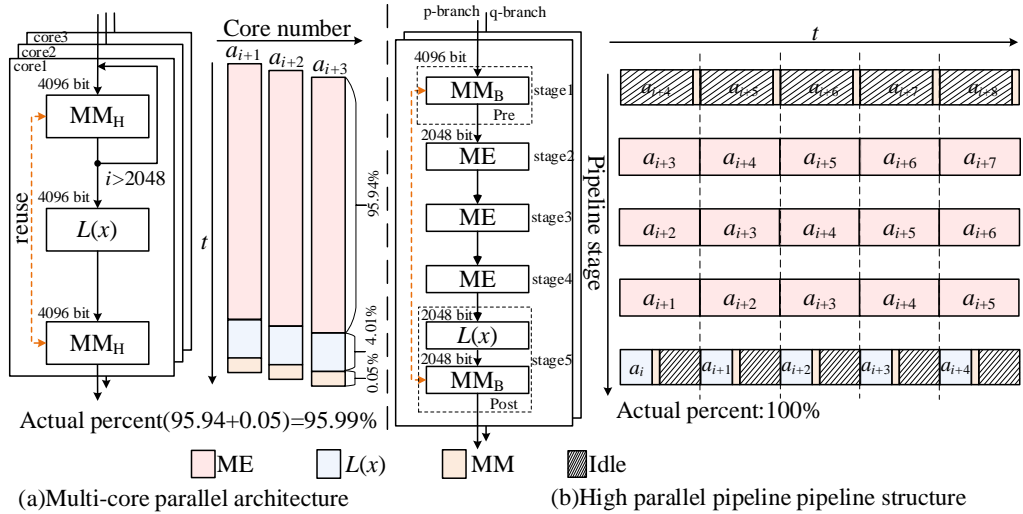

**Fig. 6.** Comparison of decryption circuit design strategies

As depicted in Fig. 6(b), we divide the decryption operation chain into preprocessing, modular exponentiation, and postprocessing, and employ a dedicated $MM_B$ unit to perform modular multiplications during preprocessing and postprocessing. Given the limited DSP resources in FPGAs, to maximize optimization for the most time-consuming ME operation, this paper allocates all DSP resources to ME while implementing $MM_B$ exclusively with LUTs. By segmenting the modular exponentiation operation into stages and organizing multiple ME units in a pipelined manner, we achieve 100% utilization of the ME units. Compared to the traditional multi-core design, in this proposed design, as the preprocessing and postprocessing units are shared by multiple MEs, the introduction of the $MM_B$ unit in preprocessing and postprocessing does not incur additional resource overhead. Furthermore, with the segmentation of the operation chain, the time for a single modular exponentiation operation in MESA can fully cover the preprocessing and postprocessing times, improving overall throughput.

## 5 IMPLEMENTATION AND COMPARISON

### 5.1 Implementation Results

This paper adopted Verilog hardware description language for hardware design and utilized Vivado 2018.3 software to experimentally validate the proposed MESA on a Xilinx Virtex-7 xc7vx690t development board. To compare with existing work, experiments were conducted under two scenarios with public key widths of 2048 and 1024 bits, respectively. Specifically, under the premise of maximizing FPGA resource utilization to achieve the highest throughput, 6 ME units (i.e., a 3-stage modular exponentiation pipeline) were deployed for the 2048-bit public key width, and 12 ME units (i.e., a 6-stage modular exponentiation pipeline) were deployed for the 1024-bit public key width. Under both parameter configurations, MESA achieved a frequency of 100 MHz. The implementation results of MESA, ME units, and pre-post processing units are summarized in Table II.

As indicated in Table II, lookup table (LUT) and register (Slice register) resources are abundant with low utilization rates. Therefore, the primary constraint of this design is the number of DSPs available. By analyzing the implementation results, we find that when $N$=2048, each ME unit consumes 526 DSPs, whereas the xc7vx690t development board only provides 3600 DSPs. Consequently, a maximum of 6 ME units can be deployed, achieving a 3-stage modular exponentiation pipeline. When $N$=1024, each ME unit consumes 270 DSPs, allowing for up to 13 ME units to be deployed. However, due to the algorithm's requirement to simultaneously compute two paths (p-branch and q-branch), and the need for an even number of ME units, a maximum of 12 ME units are used to implement a 6-stage modular exponentiation pipeline.

Table II. Implementation Result of MESA on XC7VX690T

| Module Name | | Slice LUTs | Slice Registers | DSP48E1 |
|---|---|---|---|---|
| $N$=2048 | Each ME(6) | 18384(4.24%) | 21399(2.47%) | 526(14.61%) |
| | Pre & Post | 57915(13.37%) | 77528(8.95%) | 6(0.17%) |
| | Full MESA | 168219(38.83%) | 205922(23.77%) | 3162(87.33%) |
| $N$=1024 | Each ME(12) | 9216(2.13%) | 10877(1.26%) | 270(7.5%) |
| | Pre & Post | 31049(7.17%) | 40350(4.66%) | 24(0.67%) |
| | Full MESA | 141641(32.70%) | 170874(19.72%) | 3264(90.67%) |
| Full FPGA | | 433200 | 866400 | 3600 |

Table III shows the timing results for each operation in MESA when the public key width is $N$=1024 and the number of ME units is 12. As theoretically analyzed, among the basic operators, the operations of sub, com, add, and $MM_H$ are relatively fast, taking 18, 1~18, 18, and 87 cycles, respectively. In contrast, $MM_B$ and div require significantly more time, taking 2087 and 3081 cycles, respectively. For the algorithmic steps: In the preprocessing (Pre), the $MM_B$ unit performs the Montgomery domain conversion of the ciphertext, consuming a total of 20.87 µs; The intermediate processing involves the most time-consuming modular exponentiation operation. After being divided into 6 stages, each stage of modular exponentiation takes 75.67 µs, which is also the duration of a single decryption after the pipeline is fully occupied; The postprocessing (Post) completes the decryption by performing the inverse CRT operation, taking 52.22 µs. It can be observed that the time required for a single stage of modular exponentiation fully covers the preprocessing and postprocessing stages, thus achieving full pipeline operation. As a result, under this configuration, the system can achieve 13,215 decryption operations per second.

Table III. Timing Results of each Operation when $N$=1024, No.of ME=12

| Operation | Speed | |
|---|---|---|
| | (cycles) | (μs) |
| Pre | 2087 | 20.87 |
| Each ME | 7567 | 75.67 |
| Post | 5222 | 52.22 |
| **op** | 7567 | 75.67 |
| div | 3081 | 30.81 |
| sub | 18 | 0.18 |
| com | 1~18 | 0.01~0.18 |
| add | 18 | 0.18 |
| $MM_H$ | 87 | 0.87 |
| $MM_B$ | 2087 | 20.87 |

Fig. 7 shows the relationship between performance and hardware overhead as the number of ME units increases ($N$=1024). As shown in Fig. 7(a), as the number of ME units doubles, the decryption latency is reduced by nearly half, indicating that the performance of MESA is almost doubled.

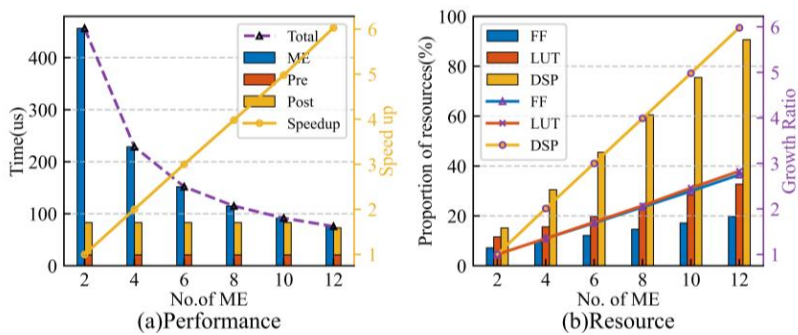

**Fig. 7. The relationship between performance, resource utilization, and the number of MEs**

However, due to the limitations of hardware resources, the number of ME units cannot be increased indefinitely. Moreover, when the number of MEs reaches 12, the time consumed by preprocessing and postprocessing is roughly equivalent to the time for one stage of modular exponentiation. At this point, the radix-2 modular multiplications ($MM_B$) and large integer division operations (div) in the preprocessing and postprocessing stages gradually become another limiting factor for increasing the number of ME units.

In this paper, instead of fully instantiating the entire decryption algorithm hardware for multi-core implementation, we only instantiate the most time-consuming ME components multiple times. As a result, as the number of ME units increases from 2 to 12 (leading to a 6-fold performance improvement), the DSP, LUT, and FF resources only increase by 6, 3, and 3 times, respectively, as shown in Fig. 7(b). This means that the proposed approach achieves a performance improvement of $N$ times with only $N/2$ times the LUT and FF resource overhead. This highlights the scalability and efficiency of the highly parallel pipeline design strategy proposed in this work.

**5.2 Comparison of Decryption Performance of Paillier**

Table IV presents the resource overhead and performance comparison of the Paillier accelerator. To ensure fairness, the comparison is conducted under the same security level. To quantify area efficiency, this paper introduces throughput per unit LUT, DSP, and FF (TP/Area) as the evaluation metric. A higher TP/Area indicates better area efficiency.

We compare the proposed MESA with [13], [14], [15], and [27] in Table IV. It can be observed that our design achieves the shortest latency and the highest throughput under both $N$=1024 and $N$=2048 bit widths. Additionally, our design achieves the best TP/LUT and TP/FF in terms of area efficiency. Specifically:

As discussed before, a Paillier encryption processor named PCP was proposed by San et al[13]. In this design, the data path within PCP is full-pipelined by exploiting parallelism between operations in Paillier, and independent operations were alternated to avoid data dependencies, thereby improving performance. However, since PCP was based on the original Paillier algorithm without algorithmic or architectural optimizations, MESA achieves throughputs that are 180.18× and 192.57× higher than PCP under two different security levels, respectively. In terms of area efficiency for LUTs and DSPs, our design also achieves improvements of at least 16.56× and 1.49× over PCP.

Table IV. Performance comparison of Paillier decryption

| Design | Platform | $N$ | Freq (MHz) | Latency (ms) | LUT/DSP/FF/BRAM | TP (kbit/s) | TP/Area | | |
|---|---|---|---|---|---|---|---|---|---|
| | | | | | | | TP/LUT | TP/DSP | TP/FF |
| San[13] | Virtex-7 | 1024 | 323 | 13.513 | 13016/27/-/74 | 74 | 5.82 | 2806.52 | - |
| Cai[14] | 65nm FDSOI | 1024 | 62.5 | 2.740 | 455000/-/-/- | 364.96 | 0.82 | - | - |
| Che[27] | S10 GX2800 | 1024 | 257 | 0.142 | 709171/461/895795/- | 7029.90 | 10.15 | **15615.22** | 8.04 |
| **MESA** | Virtex-7 | 1024 | 100 | **0.075** | 141641/3264/170874/- | **13333.33** | **96.39** | 4183.01 | **79.9** |
| San[13] | Virtex-7 | 2048 | 312 | 111.111 | 15472/27/-/- | 18 | 1.19 | 682.67 | - |
| Bahadori[15] | Zynq-7020 | 2048 | 122 | 180.72 | 1781/16/1933/5 | 11.07 | 6.36 | 708.28 | 5.86 |
| | | 2048 | 122 | 15.06 | 42871/192/53328/82 | 132.80 | 3.17 | 708.28 | 2.55 |
| Che[27] | S10 GX2800 | 2048 | 187 | 0.672 | 1213056/922/1866240/- | 2976.20 | 2.51 | **3305.45** | 1.63 |
| **MESA** | Virtex-7 | 2048 | 100 | **0.577** | 168219/3162/205922/- | **3466.20** | **21.1** | 1122.52 | **17.24** |

Throughput(TP)=($N$×1000/1024)/Latency   TP/Area=TP×1024/Area(bit/s)

Similar to our design, Cai et al. [14] also employed the Montgomery power ladder algorithm in their design to resist simple power analysis and timing attacks. To balance area and computational efficiency, they implemented a modular multiplier using a radix-256 high-radix Montgomery algorithm. However, their high-radix Montgomery algorithm retains the final judgment step, which hinders performance improvements in modular exponentiation. Additionally, due to the high radix used in the modular multiplier, the system frequency was difficult to improve, with their best system frequency listed as 62.5 MHz. Furthermore, their design was also based on the original Paillier algorithm, and under the same security level, the bit width of the decryption modular exponentiation is twice that of our design, increasing computational overhead. Compared to their design, MESA achieves a throughput improvement of 36.53× and a LUT efficiency improvement of 117.55×.

Che et al. [27] proposed a heterogeneous accelerator design for federated learning. They also deployed multiple processing elements (PEs) in an FPGA, with each PE equipped with a separate task scheduler, and implemented parallel computation using a single instruction multiple data (SIMD) approach. To reduce interactions between the FPGA and off-chip global memory, their design utilized a large amount of on-chip storage resources. As a result, our design achieves higher area efficiency for LUTs and FFs, with improvements of 9.5× and 9.94× under a 1024-bit width, and 8.41× and 10.58× under a 2048-bit width, respectively. Additionally, due to the use of more DSP units to accelerate modular multiplication, MESA achieves throughputs that are 1.9× and 1.16× higher than their design under 1024-bit and 2048-bit widths, respectively. However, MESA's DSP area efficiency is significantly lower than that of Che et al.'s design.

Bahadori et al. [15] proposed a co-design approach based on a microcode multi-core architecture to accelerate Paillier algorithm. Under a 2048-bit width, MESA achieves throughput improvements of 313.21× and 26.1× compared to their single-core and 12-core designs, respectively. In terms of area efficiency, MESA simplifies control logic design by adopting the eCRT-Paillier decryption algorithm with shortened computation chains, and the highly parallel pipelined design significantly improves hardware utilization. Additionally, redundant logic in the computation is removed to reuse design elements, greatly reducing hardware resource overhead. Compared to their single-core design, MESA achieves improvements in LUT, DSP, and FF efficiency of 3.32×, 1.58×, and 2.94×, respectively, and compared to their 12-core design, improvements of 6.66×, 1.58×, and 6.76×, respectively.

# 6 CONCLUSION

This paper proposes two methods to simplify the decryption operations for the Paillier partially homomorphic algorithm: (1) combining precomputed parameters in the post-processing, and (2) removing the judgment operations in the decryption algorithm. These methods make the control circuit lightweight and shorten the processing delay. Besides, all these modifications are theoretically proven. In terms of hardware implementation, this paper proposes a highly parallelized architecture based on exponent segmentation, which relieves the scheduling of multiple modular exponentiation units and reduces the idle time of modular exponentiation operations. With the above three methods, we designed a high-throughput and efficient Paillier accelerator named MESA. The experimental results demonstrate that our design has significant advantages over existing works in terms of performance and area efficiency.